\documentclass[runningheads]{llncs}

\usepackage{graphicx}
\usepackage{amsmath}
\usepackage{amsfonts}
\usepackage{fancyvrb}
\usepackage{xcolor}
\usepackage{listings}
\usepackage{lmodern}
\usepackage{hyperref}
\usepackage{pgfgantt}
\usepackage{float}
\usepackage{tabularx}
\usepackage{wrapfig}
\usepackage{placeins}

\usepackage{tikz}
\usetikzlibrary{positioning,arrows.meta,decorations.markings}

\newcommand{\machine}{\mathrm{mach}}
\newcommand{\sta}{\mathrm{start}}
\newcommand{\et}{\mathrm{end}}

\newcommand{\job}{\mathrm{job}}
\newcommand{\act}{\mathrm{act}}

\lstdefinelanguage{opl}{
	morekeywords={all, and, assert, boolean, constraint, constraints, cp, cplex, cumulFunc, DBConnection, DBExecute, DBRead, DBUpdate, dexpr, diff, div, dvar, else, execute, false, 
		float, floatplus, forall, from, if, in, include, infinity, int, intplus, intensity, inter, interval, invoke, key, main, max, maximize, maxint, min, minimize, mod, not, 
		optional, or, ordered, piecewise, prepare, prod, pwlFunc, range, reversed, sequence, setof, SheetConnection,
		SheetRead, SheetWrite, size, sorted, SPSSConnection, SPSSRead, stateFunc, stepFunc, stepwise, string, subject, sum, symdiff, to, true, tuple, types, union, using, with, cumulFunction},
	sensitive=true,
	morecomment=[l]{//},
	morecomment=[s]{/*}{*/},
	morestring=[b]",
}
\title{Easy, adaptable and high-quality Modelling with domain-specific Constraint Patterns}

\usepackage{graphicx}
\usepackage{amsmath}
\usepackage{amsfonts}
\usepackage{fancyvrb}
\usepackage{xcolor}
\usepackage{listings}
\usepackage{lmodern}
\usepackage{hyperref}
\usepackage{enumitem}

\usepackage{tikz}
\usetikzlibrary{positioning,arrows.meta,decorations.markings}

\begin{document}
	\title{Easy, adaptable and high-quality Modelling with domain-specific Constraint Patterns}
	\titlerunning{Domain-specific Constraint Patterns}
	%
	\author{Sophia Saller\inst{1}\orcidID{0000-0003-4817-8601} \and Jana Koehler\inst{1,2}\orcidID{0000-0001-9000-9188}}
	\authorrunning{S. Saller \and J. Koehler}
	%
	\institute{Deutsches Forschungszentrum für Künstliche Intelligenz, Saarland Informatics Campus, Saarbr\"ucken, Germany, \email{firstname.lastname@dfki.de}\and
		Universität des Saarlandes, Saarland Informatics Campus, Saarbr\"ucken, Germany}
	\maketitle              
\begin{abstract}
Domain-specific constraint patterns are introduced, which form the counterpart to design patterns in software engineering for the constraint programming setting. These patterns describe the expert knowledge and best-practice solution to recurring problems and include example implementations. We aim to reach a stage where, for common problems, the modelling process consists of simply picking the applicable patterns from a library of patterns and combining them in a model. This vastly simplifies the modelling process and makes the models simple to adapt. By making the patterns domain-specific we can further include problem-specific modelling ideas, including specific global constraints and search strategies that are known for the problem, into the pattern description. This ensures that the model we obtain from patterns is not only correct but also of high quality.
We introduce domain-specific constraint patterns on the example of job shop and flow shop, discuss their advantages and show how the occurrence of patterns can automatically be checked in an event log.
\end{abstract}
\section{Introduction}
Design patterns are widely used in software engineering to describe solutions to recurring problems. Such recurring problems do not, however, only occur in software engineering; they also appear in constraint programming. We do not want to constantly have to find new solutions to problems that have already been tackled, but would instead like to reuse methods that have worked successfully in the past. To this end, we introduce the notion of domain-specific constraint patterns: patterns for the constraint programming setting. A pattern describes a recurring problem, gives a solution approach to this problem, and highlights consequences of choosing this approach. The solution approach describes the best practice or expert knowledge of how to model the problem. By making the patterns domain-specific, problem specific modelling ideas, including global constraints and search strategies that are known for the problem, can be included into the pattern description. This allows for a simplification of the modelling process, as well as making high-quality models easier to obtain.
The domain-specific constraint patterns introduced here form an extension of the general constraint patterns introduced by Toby Walsh in \cite{walsh2003constraint}. The patterns introduced there are general modelling techniques for which no specific implementation is given. Next to introducing the notion of domain-specific constraint patterns, we discuss their potential advantages and give examples of patterns for scheduling problems by formalizing and describing them. 
Over time, we intend to build up a pattern library containing patterns from further scheduling as well as non-scheduling problems, such as timetabling, permutation or travelling salesman problems.  

\section{Why Domain-specific Constraint Patterns?}
\textbf{Easy Modelling}: Making use of domain-specific constraint patterns has many advantages. It has the potential to simplify the modelling process. We aim to reach a stage where, for common problems, the modelling process consists of a human modeller simply picking the applicable patterns from a large library of patterns and combining them in a model. Each pattern further includes code examples from which a complete model can simply be assembled once the set of patterns occurring in the problem is known. If further, less common, constraints occur in the problem, the human modeller can simply add these to the basic model structure given by the pattern implementation. This makes past solutions to recurring problems reusable. Furthermore, by either using the explicit and automatically checkable conditions included in the pattern description, or by directly checking whether the constraints are satisfied, an event log can automatically be classified according to the patterns. Informally, an event log is a collection of events, where each event corresponds to an activity that was executed. A formal definition is given in Definition~\ref{def.eventlog}. This means that, if a past event log of a problem exists, the event log can automatically be classified according to the patterns, and a constraint model can be assembled by a human modeller from the explicit implementations of the occurring patterns. This approach to acquiring a constraint model using patterns is further developed in Section~\ref{section.constrAcq}.

\textbf{Adaptable Modelling}: In real world problems the problem we wish to model may frequently change to some small extent, sometimes referred to as concept drift \cite{tsymbal2004problem}. As an example, in automotive production it may be the case that usually only one car is allowed to wait between two machines, but on busy days this might be relaxed to two or more waiting cars to allow for a greater throughput. On these occasions, the constraint model has to be adapted to fit the new situation. By specifying constraint patterns and relating the constraints directly to the patterns, all that needs to be done to adapt the model is to check which patterns no longer hold and which new patterns do, and adapt the model accordingly using the implementation specified in the patterns. This results in easy adaptations of constraint models.

\textbf{High-quality Modelling}: The use of patterns ensures that the resulting model is of high quality. For a constraint model to be useful in practice, the model needs to not only be correct, but also be of high quality to help the solver find solutions quickly.
Constraint programming experts use their vast experience and best practices when trying to model new problems in a way that simplifies the solution-finding process for the solver.
However, this expert knowledge is usually not accessible to others. Harvesting domain-specific constraint patterns makes this expert knowledge easy to obtain, which will result in it being more widely applied. In the pattern description we give explicit information on how the pattern is to be best modelled and which search strategies are known to work well for this problem. 

\textbf{Explainable Modelling}: Finally, patterns can help the explainability of the constraint model. While domain experts have an in-depth understanding of the problem at hand, they are usually not well-versed in constraint programming or the mathematical description of constraints. The proposed formalisation of each domain-specific constraint pattern contains an explanation of the intent of the pattern and examples in plain English. This makes it easier to discuss the modelling of the problem with domain-experts, since, instead of having to discuss at the abstraction level of variables and constraints, discussions can be held at the abstraction level of patterns.

\section{Related Work}
Design patterns were originally introduced in the domain of building architecture to describe best-practice or expert solutions to recurring problems \cite{alexander1977pattern}. Several years later, the idea of applying design patterns was extended to the world of programming for designing reusable solutions to common software engineering problems \cite{gamma1995design}. The importance of high-quality modelling in constraint programming has long been recognized~\cite{borrett2001context,smith2006modelling}, which emphasises the importance of making expert knowledge accessible in this setting. Constraint patterns, as design patterns for constraint programming, were first introduced by Walsh \cite{walsh2003constraint} to help constraint programmers spot commonly occurring antipatterns which make models inefficient and describe expert knowledge on how to best resolve these issues. However, the patterns introduced by Walsh are not domain-specific and instead are general modelling techniques for which no specific implementation is given. By making constraint patterns domain-specific, we allow for the inclusion of problem specific modelling ideas, which further improve the quality of the final model.

Next to resulting in high-quality models, the use of domain-specific constraint patterns also simplifies the modelling process. Other approaches to simplifying the modelling process exist in the literature as so-called constraint acquisition (CA) systems, which automatically obtain constraints from a set of examples. One has to distinguish between active and passive CA systems. Active CA systems may ask queries to the user, for example ask the user to classify an assignment as either valid or invalid. Examples of active CA systems can be found in \cite{arcangioli2016multiple,bessiere2013constraint,bessiere2017constraint,freuder2002suggestion,o2004study,tsouros2020efficient}.
Passive CA systems, on the other hand, cannot interact with the user and instead learn exclusively from a set of examples, which may consist of just positive (satisfying assignments), or negative (invalid assignments) and positive examples. Examples of passive CA systems can be found in \cite{beldiceanu2012model,bessiere2017constraint,lallouet2010learning}. CA systems for scheduling problems are introduced in \cite{kumar2019automating,senderovich2019learning}. 

In \cite{senderovich2019learning}, a passive CA system learning a constraint model from event logs, meaning from exclusively positive examples, is proposed.
In this, the authors restrict their acquisition to basic scheduling problems.
They efficiently check whether the given event log is such a problem by checking whether the Petri Net, which is mined from the event log using a process mining algorithm, is an \emph{activity resource petri net} (ARPN). 
This work proposes an end-to-end solution for a specific subclass of scheduling problems and suggest extensions to further classes of scheduling problems in future work. However, extensions to other classes are laborious as a new Petri Net class must be defined for each new problem class. For this reason, we propose an extension of their work using a more modular approach by introducing domain-specific constraint patterns. By choosing this modular approach, we additionally attempt to tackle the problem of obtaining not just valid, but good constraint models. Initial experiments of this show promising results and are introduced in Section~\ref{experiments}.
Moreover, instead of checking whether a CP model for an a priori known problem class can be constructed from an event log using a specialized algorithm, we propose declaratively formulated conditions, which can also be verified in polynomial time.
In both cases, the checking of the conditions is linear in the size of the event log. While the system introduced in \cite{senderovich2019learning} gracefully fails if the event log does not correspond to a basic scheduling problem, we can provide further explanations by returning which patterns do or do not hold and for those that do not hold, give specific events which violate the conditions of the pattern.

The proposed library of constraint patterns complements two other important libraries related to constraint problems and constraint programming, CSPlib and the global constraint catalog. 
CSPlib \cite{csplib} is an extensive library of constraint problems, which was initially published with the purpose of providing a range of hard and realistic problems that can be used to benchmark algorithms. Some of the problems in the library include example implementations in different modelling languages, such as Essence \cite{frisch2008ssence} and MiniZinc \cite{minizinc}. The goal of CSPlib is thus very different to that of the constraint pattern repository introduced here, which aims to simplify the modelling process and make high-quality models more easily obtainable. We also take one step further and try to break down the problems by analysing the structure behind them. In doing so, the domain-specific constraint patterns can easily be reused in different application scenarios. Our goal for the constraint pattern library is to eventually cover all problems introduced in the CSPlib library and provide links from patterns to relevant problems in CSPlib where applicable.
The global constraint catalog \cite{beldiceanu2005global}, on the other hand, is a collection of global constraints. A global constraint is a high-level modelling abstraction which typically aids the propagation of the constraint solver as many solvers have special, efficient inference algorithms implemented for them. For each global constraint, the global constraint catalog contains its semantic and, when available, typical usage and filtering algorithms. Constraint patterns operate on a higher modelling level compared to global constraints. As such, the modelling variants included in the description of domain-specific constraint patterns may include global constraints, but they may also include other constraints as well as information on how the variables of the problem are to be best chosen and modelled.

\section{Describing Domain-specific Constraint Patterns}
We introduce domain-specific constraint patterns on the example domain of production scheduling, which we use as a starting point to build up the pattern repository due to its practical relevance. Scheduling problems can be classified according to three characteristics: the machine environment, the processing characteristics and the objective function, using the classification schema by Graham et al. \cite{graham1979optimization}. This schema also inspired the introduction of constraint patterns, with the idea of having a pattern for each machine environment and each processing characteristic, and where the modelling of the processing characteristic pattern depends on the machine environment that occurs.
In order to illustrate the idea of a pattern library, we discuss two selected patterns, the job shop and flow shop patterns, see Tables~\ref{table.jobshop} and~\ref{table.flowshop}.
 
A domain-specific constraint pattern is a general, reusable solution to a commonly recurring problem within a given domain. In the following, we propose key elements for the description of domain-specific constraint patterns based on the examples of the Job Shop Pattern (Table~\ref{table.jobshop}) and the Flow Shop Pattern (Table~\ref{table.flowshop}). We base these key elements on the established pattern template from \cite{gamma1995design}, which we adapt to the constraint pattern setting. It includes the \emph{Intent}, explaining what the constraint pattern does, the \emph{Motivation} behind the pattern, and the \emph{Applicability} explaining in which situations the pattern can be applied. 
In the \emph{Participants} entry, the variables participating in the pattern are listed, and the \emph{Collaborations} entry explains how these participants collaborate. In the \emph{Diagram} entry, a visual representation of the pattern is given. In some cases, applying the pattern has certain \emph{Consequences} which need to be considered. The ``Implementation'' entry of the original design pattern template has been adapted for the constraint pattern setting to the new entry \emph{Modelling variants}. This entry gives variants of the best-practice modelling of the pattern for different constraint modelling languages. Inspired by~\cite{zimmermann2009managing}, the original ``See Also'' entry is refined for the constraint pattern setting to five entries: \emph{Forces} (patterns that must hold if this pattern holds), \emph{Enables} (patterns that have this pattern as a prerequisite), \emph{Incompatible with} (patterns that can not be applied with this pattern), \emph{Compatible with} (patterns that can be applied with this pattern), and \emph{Similar to} (patterns that are similar to this pattern).
Note that some of these relationships can also be directly checked. For example, if two incompatible patterns are combined in a model, the model will become unsatisfiable. Finally, a further entry is added to the template which is important in the constraint pattern setting:
The \emph{Search strategies} entry, which lists search strategies which work well in the setting of this pattern.

\medskip

In the \textit{Applicability} element of the pattern description, we give information about the pattern's applicability by giving conditions which must be satisfied by corresponding event logs. The applicability is described through necessary and sufficient conditions. Necessary conditions are other patterns which are prerequisites for the occurrence of this pattern. Sufficient conditions are further conditions that an event log must satisfy to display this pattern. We require the pattern conditions to be sound and complete. The pattern conditions are \textit{sound} if the behaviour corresponding to any event log satisfying the conditions of the pattern, also satisfies the constraints of the pattern. The pattern conditions are \textit{complete} if any pattern occurring (as sets of constraints) in a model also occurs (as satisfied conditions) in any event log obtained from taking a solution of the model and executing it. For the example patterns given in this paper, we describe the sufficient conditions based on the sequence of operations $ops(j) = \langle op_1(j), \dots\rangle$ of each job $j\in J_E$ in the event log and the sequence of operations $\mathfrak{ops}(m) = \langle \mathfrak{op}_1(m), \dots\rangle$ on each machine $m\in M_E$ in the event log. Formal definitions can be found in Definitions~\ref{def.o} and~\ref{def.a} in Section~\ref{section.constrAcq}.

\subsection{Job Shop and Flow Shop Pattern}
We give part of the description of two domain-specific constraint patterns corresponding to machine environments here: The Job Shop Pattern in Table~\ref{table.jobshop} and the Flow Shop Pattern, a specialisation of the Job Shop Pattern, in Table~\ref{table.flowshop}.
Since there are more restrictions on the operations of each job in the Flow Shop Pattern when compared to the Job Shop Pattern, the Flow Shop Pattern allows for a more streamlined implementation of the variables in the modelling variants.

The \emph{Modelling variants} entry in the pattern description specifies how to best model the pattern in different modelling languages.
In the case of the Flow Shop Pattern we give two implementation variants, one for the OPL language for the IBM ILOG CPLEX CP solver~\cite{cplex} and one for the MiniZinc language~\cite{minizinc}.
\begin{itemize}
	\item \emph{OPL language} (see Figure~\ref{fig.fs_cplex}):
	For each job, exactly $M$ interval decision variables are introduced, where $M$ is the number of machines. Further, for each job, a precedence constraint ensures that each operation of this job can only start being processed once all the predecessor operations of this job have finished processing. A sequence decision variable is introduced for each machine, containing all interval variables corresponding to operations processed on said machine. A noOverlap constraint on this sequence variable ensures that each machine only ever processes one job at a time.
	\item \emph{MiniZinc language} (see Figure~\ref{fig.fs_minizinc}): For each operation of a job an integer variable for the start time is introduced. For each job, a precedence constraint ensures that each operation of this job can only start being processed once all the predecessor operations of this job have finished processing. Further, each machine can only ever process one job at once. For this reason, a disjunctive constraint between all the operations intervals on this machine is introduced.
\end{itemize}

Note further that if several patterns hold at the same time, the implementations of the different patterns can depend on one another. Consider for example the Machine Setup Pattern, which states that each machine needs a certain amount of setup time between two operations. The Machine Setup Pattern has the Job Shop Pattern as a prerequisite. The implementation for this pattern is different depending on whether the Flow Shop Pattern also holds or not.

\begin{figure}[H]
	\centering
\begin{lstlisting}[breaklines=true,numbers=left]
int numberJobs = ...;
int numberMachines = ...;
range allJobs = 1..numberJobs;
range allMachines = 1..numberMachines; 

tuple Operation {
	string id;    // Operation id
	int jobId; // Job id
	int pos;   // Position in job
	int machine; // Machine
	int pt; // Processing time
};
{Operation} allOperations = ...;

dvar interval operations[op in allOperations] size op.pt; 
dvar sequence machineSequence[m in allMachines] in all
	(op in allOperations: op.machine == m) operations[op];
\end{lstlisting}
\begin{lstlisting}[breaklines=true,numbers=left]
forall (j in Jobs, op1 in allOperations, op2 in allOperations: op1.jobId==j &&
	op2.jobId == j && op2.pos > op1.pos)
	endBeforeStart(operations[op1],operations[op2]);

forall (m in allMachines) noOverlap(machineSequence[m]);
\end{lstlisting}
	\caption{OPL implementation of the variables and constraints for the Job Shop Pattern (Table~\ref{table.jobshop}).} \label{fig.js_cplex}
\end{figure}

\begin{table}[!ht]
	\centering
	\caption{Job Shop Pattern.}\label{table.jobshop}
	\begin{tabularx}{\textwidth}{|l|X|}
		\hline
		\multicolumn{2}{|c|}{\textbf{\rule{0pt}{2.5ex}Job Shop Pattern}}\\
		\hline
		\rule{0pt}{2.5ex}Intent & A set of jobs is to be processed on a set of machines. Each job is to be processed on some of the machines (potentially multiple times) in a predetermined order which might be different for each of the jobs. No two operations of the same job are to be processed simultaneously, and no machine is to process two operations simultaneously.\\
		\hline
		\rule{0pt}{2.5ex}Motivation & Job shop machine environments regularly occur in manufacturing scenarios such as car manufacturing.\\
		\hline
		\rule{0pt}{2.5ex}Applicability & \emph{Sufficient conditions}: For all $j\in J_E$ and all $m\in M_E$,\\
		(Necessary and& 
		(a) Jobs cannot be processed without a machine, so for all \\
		sufficient conditions)&\hspace{5mm} $i \in \{1, \dots, |ops^p(j)|\}$: $\texttt{\#}_{\machine}(op^p_i(j)) \neq \bot$. (see Def.~\ref{def.o})\\
		&(b) No two operations of a job $j$ can be processed at the same\\&\hspace{5mm} time, so for all $i \in \{1, \dots, |ops^p(j)|-1\}$:\\
		&\hspace{5mm} $\texttt{\#}_{\et}(op^p_i(j)) \leq \texttt{\#}_{\sta}(op^p_{i+1}(j))$.\\
		&(c) No machine can process two operations at the same time, so\\
		&\hspace{5mm} for all $i \in \{1,\dots, |\mathfrak{ops}^p(m)|-1\}$:\\
		&\hspace{5mm}
		$\texttt{\#}_{\et}(\mathfrak{op}^p_i(m)) \leq \texttt{\#}_{\sta}(\mathfrak{op}^p_{i+1}(m))$. (see Def.~\ref{def.a})\\
		&(d) Each activity occurs on a fixed machine, so for all $j_2 \in J$\\
		&\hspace{5mm}  and all $i_1 \in \{1, \dots, |ops^p(j)|\}$ and $i_2 \in \{1, \dots, |ops^p(j_2)|\}$:\\
		&\hspace{5mm} if $\texttt{\#}_{\act}(op^p_{i_1}(j)) = \texttt{\#}_{\act}(op^p_{i_2}(j_2))$ then\\
		&\hspace{5mm} $\texttt{\#}_{\machine}(op^p_{i_1}(j)) = \texttt{\#}_{\machine}(op^p_{i_2}(j_2))$\\
		\hline
		\rule{0pt}{2.5ex}Participants & A set of jobs consisting of operations, and a set of machines.\\
		\hline
		\rule{0pt}{2.5ex}Collaborations & Each operation of a job has to be processed on a predetermined machine. The operations of a job have to be processed in a predetermined order. The operations and route through the machines can differ between jobs.\\
		\hline
		\rule{0pt}{2.5ex}Diagram & Gantt chart showing four jobs being processed on three machines:\\
		&
		\includegraphics[width=6cm]{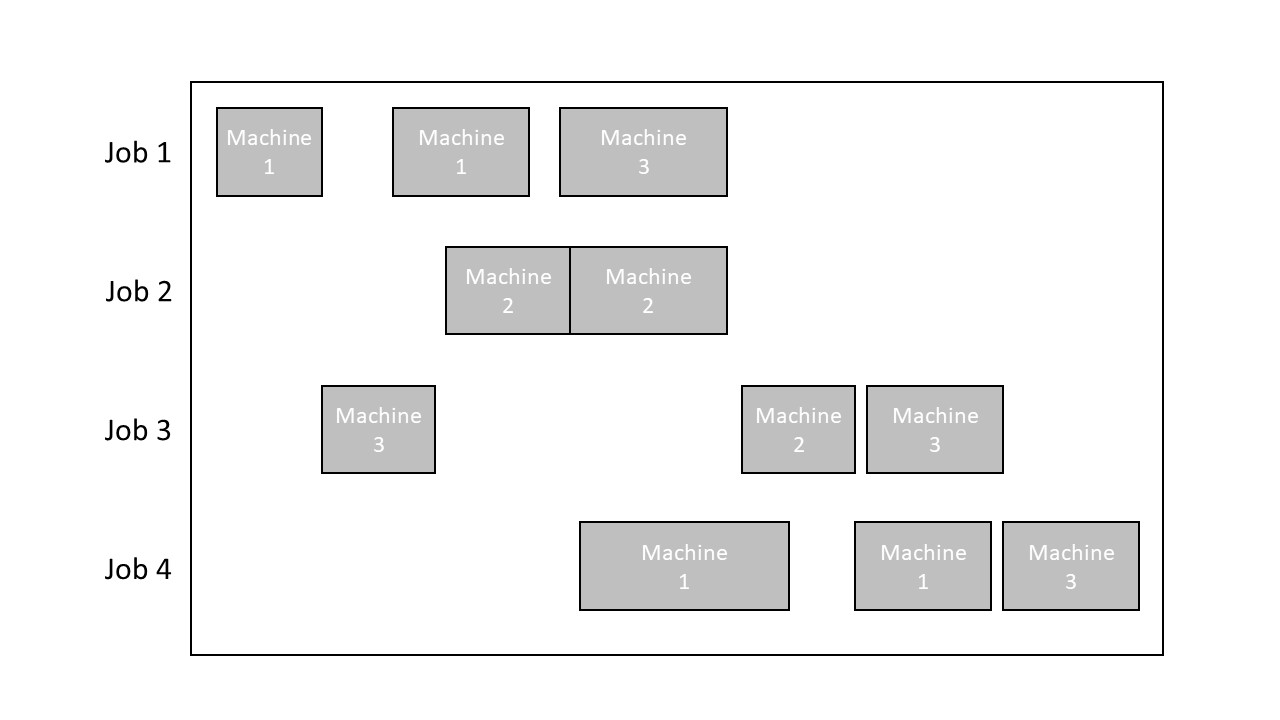}\\
		\hline
		\rule{0pt}{2.5ex}$\vdots$ & $\vdots$ \\
		\hline
		\rule{0pt}{2.5ex}Modelling Variants & \emph{(1) OPL for IBM ILOG CPLEX CP}: See Figure~\ref{fig.js_cplex}, $\dots$\\
		\hline
		\rule{0pt}{2.5ex}$\vdots$ & $\vdots$ \\
		\hline
		\rule{0pt}{2.5ex}Search strategies & \textit{Makespan optimisation}: Set start times by choosing the job that can start earliest and setting it to that time. If the total end time is not fixed, we set it to its minimal possible value~\cite[Section 2.5.2.]{MiniZincDoc}.\\
		\hline
	\end{tabularx}
\end{table}

\FloatBarrier

\begin{table}[!ht]
	\centering
	\caption{Flow Shop Pattern.}\label{table.flowshop}
	\begin{tabularx}{\textwidth}{|l|X|}
		\hline
		\multicolumn{2}{|c|}{\rule{0pt}{2.5ex}\textbf{Flow Shop Pattern}}\\
		\hline
		\rule{0pt}{2.5ex}Intent & A set of jobs is to be processed on a set of machines. Each job is to be processed on each of the machines exactly once in a predetermined order and the order is the same for all jobs.\\
		\hline
		\rule{0pt}{2.5ex}$\vdots$ & $\vdots$ \\
		\hline
		\rule{0pt}{2.5ex}Applicability & \textit{Necessary conditions}: Job Shop Pattern (Table~\ref{table.jobshop}).\\
		&\textit{Sufficient conditions}: ...\\
		\hline
		\rule{0pt}{2.5ex}$\vdots$ & $\vdots$ \\
		\hline
		\rule{0pt}{2.5ex}Consequences & The decision variables define that each job consist of exactly as many operations as there are machines.\\
		\hline
		\rule{0pt}{2.5ex}Forces & Job Shop Pattern (Table~\ref{table.jobshop}), ...\\
		\hline
		\rule{0pt}{2.5ex}Enables & No Wait Pattern, ...\\
		\hline
		\rule{0pt}{2.5ex}Compatible with & Distinguishable Resources Pattern, Indistinguishable Resources Pattern, ...\\
		\hline
	\end{tabularx}
\end{table}

\FloatBarrier

\begin{figure}[!ht]
	\centering
\begin{lstlisting}[breaklines=true,numbers=left]
include "disjunctive.mzn";
int: horizon;  % time horizon
set of int: Time = 0..horizon;
int: numMachines;
int: numJobs;

array[1..numJobs, 1..numMachines] of int: productionTimeMatrix;
int: maxd = max([productionTimeMatrix[j,m] | j in 1..numJobs, m in 1..numMachines]);

array[1..numJobs, 1..numMachines] of var Time : s;
array[1..numJobs, 1..numMachines] of 0..maxd : d = productionTimeMatrix;  
\end{lstlisting}
\begin{lstlisting}[breaklines=true,numbers=left]
constraint
forall(j in 1..numJobs, m1 in 1..numMachines, m2 in m1+1..numMachines)(
    d[j, m1] <= s[j, m2] ~- s[j, m1]);

constraint
forall(m in 1..numMachines)(
  disjunctive(s[1..numJobs, m], d[1..numJobs, m]));
\end{lstlisting}
	\caption{MiniZinc implementation of the variables and constraints of the Flow Shop Pattern.} \label{fig.fs_minizinc}
\end{figure}
\begin{figure}[H]
	\centering
\begin{lstlisting}[breaklines=true,numbers=left]
int operationDurations[allJobs][allMachines] = ...;
dvar interval operations[j in allJobs][m in allMachines] 
	size operationDuration[j][m];
dvar sequence machineSequence[m in allMachines] in all(j in allJobs)
	operations[j][m];
\end{lstlisting}
\begin{lstlisting}[breaklines=true,numbers=left]
forall(j in allJobs, m1 in allMachines, m2 in allMachines: m1<m2)
    endBeforeStart(operations[j][m1], operations[j][m2]);
    
forall(m in allMachines) noOverlap(machineSequence[m]);
\end{lstlisting}
	\caption{OPL implementation of the variables and constraints of the Flow Shop Pattern.} \label{fig.fs_cplex}
\end{figure}

\FloatBarrier

\section{Automatic pattern detection in event logs: Towards a DevOps Chain for Constraint Programming}\label{section.constrAcq}
As part of the description of patterns, we give information about the applicability of the pattern, meaning the situation in which the pattern can be applied. We describe these conditions as necessary and sufficient conditions on an event log. To formally state these, we make some definitions for event logs below using the formalisation by Wil van der Aalst in \cite{van2016data}. To automatically check the presence of patterns in an event log, the event log needs to be brought into this standard form.
Note that we do not make any assumptions on the quality of the event log with respect to the objective function or about the structure of the process or problem class to which the process  described by the event log belongs.

We now define what we mean by an event and an event log, which may be given in the XES Standard for event logs \cite{xes2016ieee}. Table~\ref{table.example} shows a small example of a scheduling event log. \footnote{Further examples of event logs, including logs from non-scheduling domains, are available at \url{www.processmining.org/event-data.html}.}
\medskip

\begin{definition}[Event, attribute]\cite{van2016data}
	Let $\mathcal{E}$ be the event universe. Events may be characterized by various attributes. Let $AN$ be the set of attribute names. For any event $e\in\mathcal{E}$ and name $n\in AN$, $\texttt{\#}_n(e)$ is the value of attribute $n$ for event $e$. If event $e$ does not have an attribute named $n$, then $\texttt{\#}_n(e) = \bot$ (null value).
\end{definition}
We can now define an event log. An event log contains data related to a single process and consists of events. Each event refers to a single process instance, also referred to as case. For a set $A$, let $A^*$ denote the set of finite sequences over $A$.
\begin{definition}[Event Log $E$]\cite{van2016data}
	\label{def.eventlog}
	Let $\mathcal{C}$ be the case universe. For any case $c\in \mathcal{C}$ and name $n\in AN$, define $\texttt{\#}_n(c)$ to be the value of attribute $n$ for case $c$ ($\texttt{\#}_n(c) = \bot$ if case $c$ has not attribute $n$). Each case has a special attribute trace, where $\texttt{\#}_{\mbox{trace}}(c) = \mathcal{E}^*$. A trace is a finite sequence of events such that each event appears only once.
	An event log is then a set of cases $E\subseteq \mathcal{C}$ such that each event appears at most once in the entire log.
\end{definition}
Events that belong to the same activity instance, for example the start processing and complete processing event of one activity, may be identified as belonging together through the activity instance attribute or the activity name.
We also call an activity instance an operation. In the following, we will restrict our attention to the attributes necessary to describe the applicability of the patterns introduced in this paper.
\begin{table}[!ht]
	\centering
	\caption{Example of an event log with jobs as cases.}\label{table.example}
	\begin{tabular}{|c|c|c|c|c|c|c|}
		\hline
		\rule{0pt}{2.5ex}\hspace{1mm}Case\hspace{1mm} & \hspace{1mm}Event\hspace{1mm} & Timestamp & Activity & \hspace{1mm}Transaction\hspace{1mm} & \hspace{1mm}Machine\hspace{1mm} & \hspace{1mm}Activity\hspace{1mm}\\
		($j$) &&&&&($m$)&instance\\
		\hline
		$j1$ & 1 & \hspace{1mm}30-12-2010:00.00:00\hspace{1mm} & \rule{0pt}{2.5ex}\hspace{1mm}heating\hspace{1mm} & start processing & $m1$ & 1\\
		 & 2 &  30-12-2010:00.01:30 & heating & complete processing & $m1$ & 1\\
		 & 5 &  30-12-2010:00.02:00 & rolling & start processing & $m2$ & 2\\
		 & 6 &  30-12-2010:00.10:00 & rolling & complete processing & $m2$ & 2\\
		$j2$ & 3 &  30-12-2010:00.01:30 & heating & start processing & $m1$ & 3\\
		 & 4 &  30-12-2010:00.03:30 & heating & complete processing & $m1$ & 3\\
		 & 7 &  30-12-2010:00.11:10 & rolling & start processing & $m2$ & 4\\
		 & 8 &  30-12-2010:00.14:00 & rolling & complete processing & $m2$ & 4\\
		\hline
	\end{tabular}
\end{table}
Note that the same process might be described by different event logs. For the example of scheduling problems, an event log might be using the jobs as cases, or the machines as cases.
While some events will occur in both logs, other events such as machine setup will only occur in one of the logs. From an event log, we can obtain, by matching all events of the same activity instance, a sequence of operations for each job and each machine. If a pattern has sufficient conditions stated over for example the machine sequences, but they are empty in this case, the pattern is clearly not satisfied. These sequences can be obtained even if the log looks different, for example if it contains start time and duration information of each operation instead of start and completion time. In these cases, a suitable adapter needs to be implemented for the given event log structure.
\begin{definition}[Job operation sequence $ops(j)$]
	\label{def.o}
	For each job $j \in J_E$ in the event log $E$, the \emph{job operation sequence} $ops(j) = \langle op_1(j), \dots, op_{|ops(j)|}(j)\rangle$ is the sequence of operations corresponding to grouped events of the trace of the job ordered by ascending start times (if these exist), where each operation $op_{i}(j)$ may have any number of attributes. Attributes can for example be the following: the activity $\texttt{\#}_{\act}(op_i(j))$, the start time $\texttt{\#}_{\sta}(op_i(j))$, the completion time $\texttt{\#}_{\et}(op_i(j))$ and the machine $\texttt{\#}_{\machine}(op_i(j))$ of the operation.
\end{definition}

For example, taking the event log from Table~\ref{table.example}, $op_1(j1)$ corresponds to the first two rows in the table, meaning the start time is $\texttt{\#}_{\sta}(op_1(j1)) =$ 30-12-2010:00.00:00, the end time is $\texttt{\#}_{\et}(op_1(j1)) =$ 30-12-2010:00.01:30 and the machine is $\texttt{\#}_{\machine}(op_1(j1)) = m1$.
We define operation sequences for machines analogously to the operation sequences for jobs.\begin{definition}[Machine operation sequence $\mathfrak{ops}(m)$]
	\label{def.a}
	For each machine $m \in M_E$, the \emph{machine operation sequence} $\mathfrak{ops}(m) = \langle \mathfrak{op}_1(m), \dots, \mathfrak{op}_{|\mathfrak{ops}(m)|}(m)\rangle$ is the sequence of operations corresponding to grouped events of the trace of $m$ ordered by ascending start times (if these exist), where each operation $\mathfrak{op}_{i}(m)$ may have any number of attributes. Attributes can for example be the following: the activity $\texttt{\#}_{\act}(\mathfrak{op}_i(m))$, the start time $\texttt{\#}_{\sta}(\mathfrak{op}_i(m))$, the completion time $\texttt{\#}_{\et}(\mathfrak{op}_i(m))$, and the job $\texttt{\#}_{\job}(\mathfrak{op}_i(m))$ of the operation.
\end{definition}
Note that, as for events and cases, we use the notation $\bot$ to denote that an operation does not have an attribute. Further, we may restrict these operation sequences to events of a certain type, such as processing events, where the transaction type is either start processing or end processing, also called filtering. We denote the processing subsequences of $ops(j)$ and $\mathfrak{ops}(m)$ by $ops^p(j)$ and $\mathfrak{ops}^p(m)$ respectively.

Note that event logs often contain some amount of error, caused for example by noise in the sensory data. Since we have stated the conditions separately from the constraints, we can allow for small margins of errors in the conditions without affecting the pattern or constraints. This is the main reason why we separate conditions from the implementations and do not directly check the constraints, which is, of course, possible, too. As an example, we could allow for a margin of error $\delta$ in condition (b) of the Job Shop Pattern in Table~\ref{table.jobshop}:
Let $j_1\in J_E$. For all $j_2\in J_E$ and all $i\in\{1,\dots,|ops^p(j_2)|\}$ we have
\[\texttt{\#}_{\et}(op^p_i(j_2)) \leq \texttt{\#}_{\sta}(op^p_{i+1}(j_2)) + \delta.\]

\medskip

By using the conditions on the event logs together with the explicit implementations of the patterns, we can detect which patterns occur in an event log and then assemble a model from the explicit implementations of the occurring patterns. 
In reality, though, the obtained model does not remain statically true, but often needs to be revised over time. There can be several reasons for this. For example, external events may have impacted the problem and thus modified its characteristics, which is sometimes referred to as problem drift or concept drift \cite{tsymbal2004problem}, or the original model did not correctly capture all aspects of the problem. It is therefore important to keep observing the execution of the solutions to correct, improve or adapt the model.
We can do so by detecting changes in the patterns that are currently present, by either checking the conditions given in the description of the patterns, or by directly checking if the corresponding set of constraints is satisfied.

We wish to achieve a stage where when and how the problem changes, and how the model should change to fit the
current situation, is continuously automatically
detected. 
While the detection of changes in the occurring patterns is to be automated, the new model is to be assembled from the explicit implementations in the patterns by a human modeller, who reuses expertise available in the patterns. However, one can also arrive at a fully automatic solution similar to the one proposed by Senderovic et al. in \cite{senderovich2019learning} when the problem falls into a designated problem class, for which the configuration and parametrization of the patterns is straightforward. 

\begin{figure}[H]
	\begin{center}
		\includegraphics[width=7cm]{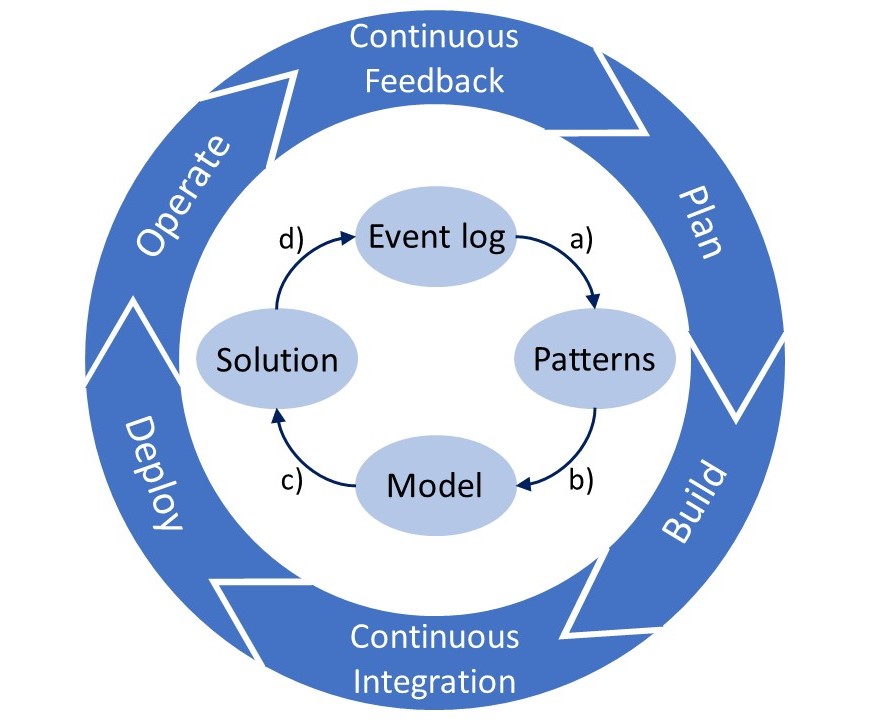}
	\end{center}
	\caption{Workflow between event log, patterns, constraint model, solution and execution, consisting of the following four steps: a)~Automatically detecting which patterns occur in the problem. b)~Assembling a parametrised constraint model from the patterns by using the explicit implementations of the patterns. c)~Finding solutions to future instances of the problem by using the model and a constraint solver. d)~Executing a solution schedule to again obtain an event log.} \label{fig.circdependencies}
\end{figure}

%
We describe this agile constraint modelling and deployment as CPDevOps, the \mbox{DevOps} for constraint programming, leading to the circular dependencies shown in Figure~\ref{fig.circdependencies}.
A similar loop was proposed in~\cite{bessiere2017inductive} as the inductive constraint programming loop. In this the main focus is however on acquiring values of parameters from past data, rather than new or changing constraint patterns. This continuous estimation of parameter values should also be included in the CPDevOps cycle to further automate the process of continuous adaption and improvement of the constraint model to capture any drift in the problem. 

\section{Experiments}\label{experiments}
In~\cite{senderovich2019learning}, the authors evaluate their constraint model acquisition system on two domains: simulated event logs generated using publicly available Job Shop benchmarks and real-world event logs from an outpatient cancer hospital in the United States. While the second data set is not publicly available, the first is available under \url{https://github.com/mnmlist/JobShopScheduling}. In the following, we compare the performance of the model acquired from the event log using the method proposed in~\cite{senderovich2019learning}, referred to here as ARPN Model, with the model obtained using domain-specific constraint patterns, referred to here as Pattern Model, on this Job Shop data set. As no information about the hardware used for the experiments is given in~\cite{senderovich2019learning}, we ran all the experiments using the ARPN model again to make the results comparable. We were in touch with the authors to obtain their original model, however did not receive their model in time. The ARPN Model was thus assembled using the model description available in~\cite{senderovich2019learning}. 

The dataset comprises 53 job shop instances. Each instance file contains information about the jobs to be scheduled, the machines assignments for each operation, the order of the operations within a job, and the operation durations. The number of jobs in an instance ranges from 6 to 30 and the number of machines from 5 to 15.

The CP model experiments were run in the IDE of IBM ILOG CPLEX Optimization Studio using an OPL script, on a Windows 10 machine with four 1.99 GHz processors and 16 GB memory. Further, the default solver settings were used for all experiments. As in~\cite{senderovich2019learning}, we use Version 12.8 of IBM ILOG CPLEX and a 10-minute time limit was set for the experiments. We performed several runs of the experiments and noticed only very minor differences in the runtime and solution costs between the runs. We thus felt that presenting numbers from a single run gives a better impression of the results than the average from several runs.

In our experiments, the solver could solve all instances with both models. Using the Pattern Model, 49 of the 53 instances could be solved to optimality compared to 48 instances using the ARPN Model. Further, all instances that were solved to optimality with both models, were solved quicker using the Pattern Model compared to the ARPN Model. On average (across all instances solved to optimality using both models) the optimal solution was found after 8.4655s using the Pattern Model, and after 21.4560s using the ARPN Model. The difference in performance is even more significant when looking at instances that could not be solved to optimality in less than 30s using either model, but were solved to optimality within the time limit. Across these 5 more difficult instances, the optimal solution was found on average after 49.08s using the Pattern Model, and after 135.58s using the ARPN Model. An overview of the quantiles of the solution times is given in Table~\ref{table.opt}. We further ran experiments with a 2min and a 5min time limit on the five instances that could not be solved to optimality using the ARPN model. As can be seen in Table~\ref{table.suboptimal}, the solver converges to good solutions only slightly quicker using the Pattern Model when compared to the ARPN model and after 10min only small differences in the solution quality can be seen. 

\begin{table}[H]
	\centering
	\caption{Solving time of instances solved optimally using both models.}\label{table.opt}
	\begin{tabular}{|c|c|c|}
		\hline
		\rule{0pt}{2.5ex}\hspace{1mm}\hspace{1mm} & \hspace{1mm}Pattern Model\hspace{1mm} & \hspace{1mm}ARPN Model\hspace{1mm}\\
		\hline
		\rule{0pt}{2.5ex}$1^{\text{st}}$ Quantile & 0.1725s & 0.2185s \\
		$2^{\text{nd}}$ Quantile & 2.0450s & 2.9695s\\
		$3^{\text{rd}}$ Quantile & 6.4545s & 13.6168s\\
		\hline
	\end{tabular}
\end{table}

\begin{table}[H]
	\centering
	\caption{Objectives of instances solved suboptimally using both models.}\label{table.suboptimal}
	\begin{tabular}{|c|c|c|c|c|c|c|}
		\hline
		\rule{0pt}{2.5ex}\hspace{1mm}Model\hspace{1mm} & \multicolumn{3}{c|}{Pattern Model} & \multicolumn{3}{c|}{ARPN Model}\\
		&\hspace{1mm}2min\hspace{1mm} & \hspace{1mm}5min\hspace{1mm} & \hspace{1mm}10min\hspace{1mm} & \hspace{1mm}2min\hspace{1mm} & \hspace{1mm}5min\hspace{1mm} &\hspace{1mm}10min\hspace{1mm}\\
		\hline
		\rule{0pt}{2.5ex}abz7 & 670 & 670 & 670 & 679 & 677 &676\\
		abz8 & 682 & 679 & 674 & 683 & 681 & 681\\
		abz9 & 696 & 696 & 696 & 701 & 689 & 686\\
		la29 & 1164 & 1164 & 1164 & 1170 & 1167 & 1165\\
		la38 & 1201 & 1201 & 1196 & 1201 & 1201 & 1196 \\
		\hline
	\end{tabular}
\end{table}
\vspace{-1cm}
\section{Conclusion and Future Work}
We introduce the idea of a repository of domain-specific constraint patterns that can automatically be checked, even on noisy event logs of processes, without a priori knowledge of the problem class into which a process falls. The patterns satisfied by an event log can be easily assembled and parametrised by human expertise, or a fully automatic construction for specific problem classes and pattern combinations, to obtain high-quality constraint models. Finally, CPDevOps, a DevOps for constraint programming, was introduced to automatically detect the constraint patterns in event logs. This enables the continuous recognition of any problem drift and the adaption of the model to fit this drift. An online pattern library is available at \url{www.constraintpatterns.com} and open for contributions by the constraint programming community of researchers and practitioners.

\vspace{-.3cm}
\bibliographystyle{splncs04}
\bibliography{lit}

\end{document}